\documentclass{article}
\usepackage[utf8]{inputenc}
\usepackage{fullpage}
\usepackage{graphicx}
\usepackage{caption}
\usepackage{float}
\usepackage{booktabs}
\usepackage{siunitx}
\usepackage{wrapfig}
\usepackage{hepnames}
\usepackage{tikz}
\usepackage[]{tikz-feynman}
\usepackage[style=nature]{biblatex}
\title{Can a muon collider be operational within the next 30 years?}
\author{Rebecca Taylor \\ University of Birmingham, School of Physics \& Astronomy \\ Physics Critique}
\date{January 2020 \\ \vspace{0.5ex} \small{Word count: 5475}}

\begin{document}

\maketitle

\begin{abstract}
    Muon colliders are a proposed next-generation particle accelerator which benefit from the muon's fundamentality and relatively high mass to perform simultaneous high precision, high energy experiments. This critique reviews their physics potential and technological feasibility, then proposes a roadmap for how a \SI{3}{\tera\electronvolt} muon collider could be built, concluding that it is experimentally possible for a muon collider to be operational in 30 years.
\end{abstract}

\section{Introduction}
This critical review will assess available evidence to answer the question of whether muon colliders can be operational by 2050.\\
Decisions on the next high energy accelerator after the Large Hadron Collider (LHC) are being considered by the high energy physics community, facilitated by the European Particle Physics Strategy Update 2020. Proposed accelerator designs were presented at the Open Symposium in May 2019, and by May 2020, the European Strategy Group (ESG) will have reviewed these proposals and will provide advice on which experiments should be considered priority for further research and development (R\&D) over the next 7 years. This strategy group is composed of scientific delegates from major European laboratories and representatives from CERN member states.\\

This review will take the position of the ESG to look at available evidence on recent progress of muon colliders, such as that presented within Section 10.4 of the ESG briefbook~\cite{Richard:BriefBook}. Other next-generation accelerators proposed in the briefbook, such as FCC, a circular electron collider, and CLIC, a linear electron collider, are focusing on producing Higgs bosons in order to measure the Higgs resonance peak. This review will conclude that the muon collider has better potential in measuring the Higgs resonance than these options and functions as a high energy discovery machine in addition. After this, the major technological components of a muon collider will be reviewed, to determine if they can be available within 30 years. Sufficient physics motivation and technological availability is required to conclude that muon colliders can be operational by 2050, and this conclusion assumes there are no significant financial or political difficulties.
\section{Motivation}
\label{sec:motivation}
The motivations and physics benefits of a muon collider must justify the high costs associated with the R\&D and technological difficulty. The major motivations for building a muon collider have been summarised by the Muon Collider Working Group, who produced a short document contributing to the ESG~\cite{Delahaye:Strategy}. There are two properties of a muon which provide these benefits: their point-like nature, and their mass of \SI{105.66}{\mega\electronvolt}, which is heavy relative to an electron \cite{Tanabashi:PDG}.

\subsection{Comparisons with other colliders}
\subsubsection{Hadron colliders}
\begin{figure}[h!]
    \centering
    \includegraphics[width=0.5\textwidth]{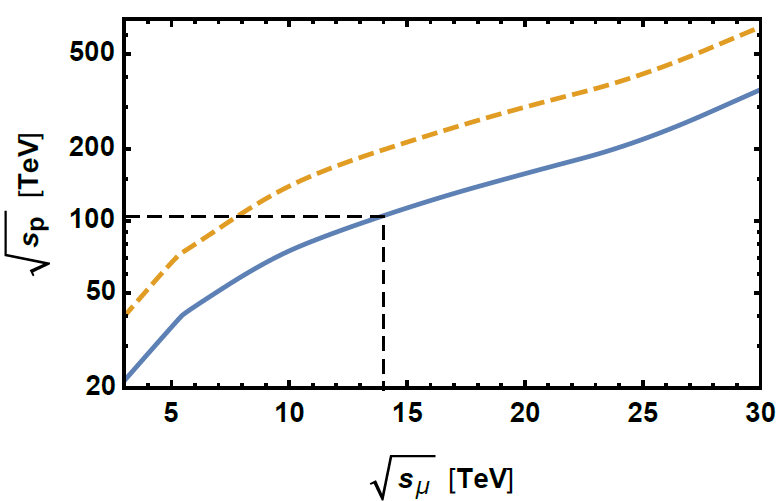}
    \caption{Equivalent proton $\sqrt{s}$, plotted as a function of muon $\sqrt{s}$ for the same collision cross-sections. Dashed orange line assumes production processes for muons and protons have roughly the same Feynman amplitudes, while the solid blue line factors in QCD effects. \cite{Delahaye:Strategy}}
    \label{fig:proton}
\end{figure}

When hadrons collide, only use a fraction of their energy is converted, due to the individually colliding quarks and gluons (partons) making up less than a third of the proton's mass. Muons are fundamental, so the entirety of its kinetic energy can be used within the collision. Therefore protons require a higher center of mass energy than muons to get the same equivalent cross-section. \\ The solid blue line in Figure~\ref{fig:proton} shows a comparison of muon center of mass energy to the proton center of mass energy, with a \SI{14}{\tera\electronvolt} muon collider having an equivalent collision cross-section to the of the proposed \SI{100}{\tera\electronvolt} FCC-hh accelerator. This demonstrates that muon colliders would be an advantageous high energy discovery machine, and could have a physics reach competitive with FCC-hh. \\

The interactions of quarks and gluons in hadronic collisions causes a high pile-up of many events, producing many jets of particles. The fundamentality of muons leads to a cleaner environment than protons, equivalent to electron collisions, making it easier to trigger on significant events. \\In addition, as partons have an unequal energy distribution, they can scan over a range of energies, whereas muons are approximately monoenergetic, within a small energy distribution. This can give them a good resolution to measure resonance peaks. \\Both of these effects together allow muon colliders to act as a precision machine.

\subsubsection{Electron colliders}
Electron colliders are limited by the emitted bremsstrahlung radiation at high energies in a synchrotron, therefore high energy electron collisions are only feasible via a large radius ring (FCC), or using linear colliders (CLIC). The amount of energy loss per synchrotron orbit ($\Delta$E) due to bremsstrahlung radiation is proportional to particle energy (E), mass (m) and synchrotron radius (r) according to Equation \ref{eq:brem} (Rearranged from the Larmor formula).
\begin{equation}
    \label{eq:brem}
    \Delta E \propto \frac{E^4}{m^4 r}
\end{equation}
As the muon-electron mass ratio is $\frac{m_\mu}{m_e} = 206.76$, this means the bremsstrahlung loss per orbit for a same radius ring is \num{5.5E-10} times smaller for muons than electrons. Therefore unlike electrons, muons will not emit a significant amount of bremsstrahlung at the TeV scale. \\

Muon colliders have a better luminosity scaling compared to electron colliders. Luminosity is a measure of brightness of the beam, which depends on its area and number of particles. A higher luminosity means more collisions, and an increased rate of events. Circular electron accelerators lose luminosity due to bremsstrahlung, and linear electron accelerators have luminosity dependant on the input RF power, resulting in a high power-consumption, on the scale of mega-watts. The instantaneous luminosity, $\mathcal{L}$, of muons is given in Equation \ref{eq:lumi}, and increases as the relativistic factor, $\gamma$, squared, effectively the energy squared. This is for magnetic field $B$, number of particles in bunch $N_0$, emittance $\epsilon$ and frequency $f_r$. (Derivation from \cite{Schulte:lumi}, European Physical Society Conference on High Energy Physics, Slides 32-26.)
\begin{equation}
\label{eq:lumi}
  \mathcal{L} \propto B \frac{N_0^2}{\epsilon \epsilon_L} f_r \gamma^2
\end{equation}

\begin{figure}[!h]
    \centering
    \includegraphics[width=0.7\textwidth]{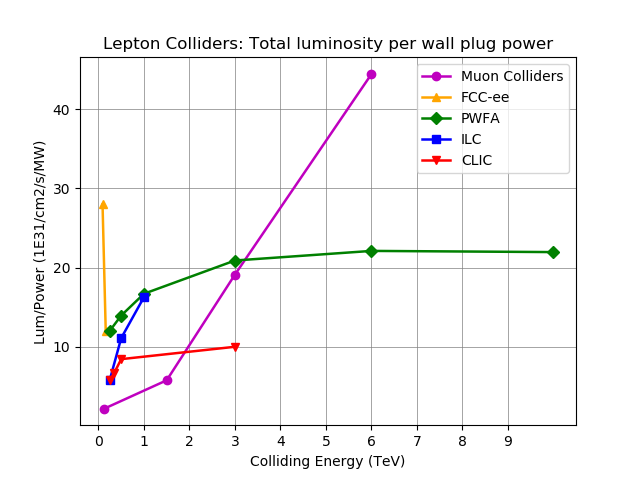}
    \caption{Luminosity scaling (\si{\centi\meter\squared\per\second}) per unit power (MW) for muon colliders compared to other proposed accelerators. Data from J.P. Delahaye \cite{Boscolo:FutureMuon}}
    \label{fig:plot}
\end{figure}

 Figure \ref{fig:plot} compares luminosity per wall plug power of proposed accelerators as a function of the center-of-mass energy \cite{Delahaye:staged}. Muon colliders in pink are shown to have the maximum luminosity per unit of input power. Linear electron colliders are CLIC in red and the ILC in blue, the circular collider FCC-ee is in yellow and plasma wakefield accelerators in green. It can be concluded that muons are the most efficient at high energies for providing good luminosity without significant power usage.

\vspace{-1em}
\subsection{Physical challenges}
\label{sec:challenges}
The physical and experimental challenges of designing and building a muon collider are due to the muon's short lifetime. They decay $\approx$ 100\% of the time via the $\Pmuon \rightarrow \Pelectron + \APnue + \Pnum $ channel \cite{Tanabashi:PDG}, with an average lifetime of $\tau_0=$ \SI{2.197}{\micro\second} at rest, which increases relativistically due to time dilation. The production, acceleration and collision of the beam must all occur on this scale, which means it is not possible to use a design which will store them for a long time, such as the LHC. Furthermore, it is experimentally challenging to get ideal beam conditions within this time, i.e. significantly reducing the emittance to improve the luminosity. This is the issue behind muon beam cooling, which is discussed in Section \ref{sec:cooling}. \\

One further difficulty the decay causes is the beam radiation background. The electrons will affect the background at the collision point of the detector and the emitted neutrino radiation may become so high intensity that it poses a serious radiation risk through high energy scattered secondary particles. These effects will be discussed further in Section \ref{sec:detneut}.

\subsection{Physics}
\label{sec:physics}
\subsubsection{Higgs physics}

Following the 2012 discovery of the Higgs Boson, C.~Rubbia presented the case for a Higgs Factory from the collisions of cooled muon beams~\cite{rubbia:Higgs}.

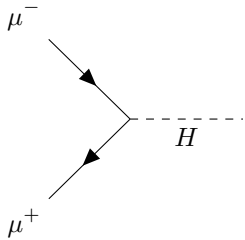
\begin{figure}[h!]
\centering
\begin{tikzpicture}
    \begin{feynman}
    \vertex (b);
    \vertex [above left=of b] (a) {\(\mu^{-}\)};
    \vertex [below left=of b] (f1) {\(\mu^{+}\)};
    \vertex [right=of b] (c);
    
    \diagram* {(a) -- [fermion] (b) -- [fermion] (f1),
      (b) -- [scalar, edge label'=\(H\)] (c),
    
    };
    \end{feynman}
\end{tikzpicture}
    \caption{s-channel Higgs production from \Pmuon \APmuon annihilation}
    \label{fig:muonchannel}
\end{figure}

The $\Pmuon \APmuon$ collision has an s-channel directly to the Higgs, shown in Figure \ref{fig:muonchannel}, with a cross-section, corresponding to Equation \ref{eq:HiggsCrossSec}~\cite{Barger:physics1998}. In addition the muon beams have a low energy spread, which is predicted to be $\frac{\sigma_E}{E} = 0.1\%$ for most muon collider designs, but could be as low as 0.004\% for a Higgs factory muon collider~\cite{Delahaye:Strategy}. The s-channel resonance can be used for measurement of the Higgs partial width.\\

\begin{equation}
    \sigma_{\Ph}= \frac{4\pi\Gamma(\Ph\rightarrow\Pmuon\APmuon)\Gamma(\Ph\rightarrow X)}{(s-m_{\Ph}^2)^2 + m_{\Ph}^2} 
    \label{eq:HiggsCrossSec}
\end{equation}

As the Higgs boson couples directly to mass, the cross-section of Higgs production scales as $m^2$ of colliding particles. The cross-section improvement of colliding muons compared to electrons is shown in Equation \ref{eq:mumu-ee} \cite{Palmer:Higgs}. The ratio of the muon and electron masses squared is \num{4.28E4}, which can provide a cross-section larger than the Higgstrahlung cross-section from electrons, $\sigma (\APelectron \Pelectron \rightarrow \PHiggs \PZ)$. \\

\begin{equation}
    \sigma(\APmuon \Pmuon \rightarrow \PHiggs) = ( {\frac{m_{\Pmu}}{m_{\Pe}}} )^2  \times \sigma (\APelectron \Pelectron \rightarrow \PHiggs)
    \label{eq:mumu-ee}
\end{equation}

The improved resolution and cross-section means muon colliders could differentiate between standard model Higgs, or any additional theoretically proposed Higgs, such as supersymmetry candidates (SUSY). In addition, if the HL-LHC upgrade is not able to observe the low branching ratio of $\PHiggs \rightarrow \Pmuon \APmuon$, then the muon collider Higgs factory is likely to measure this. Like gluon-gluon fusion at the LHC, muon-colliders can also provide a direct measurement of the top quark–Higgs Yukawa coupling via $\Pmuon + \APmuon \rightarrow \Ptop \APtop \PHiggs$.

\subsubsection{High energy physics}
\begin{figure}[h!]
    \centering
    \includegraphics[width=0.25\textwidth]{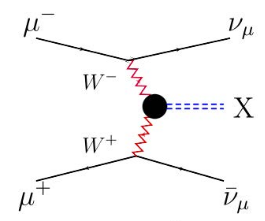}
    \caption{Feynman diagram of particle X production from vector boson fusion}
    \label{fig:boson}
\end{figure}

In 2013, the Muon Accelerator Program (MAP) submitted a contribution to Snowmass, the US equivalent of the ESG, with Section 2.4.1 dedicated to physics at muon colliders~\cite{delahaye:SNOWMASS}, including possible high-energy discoveries at the \si{\tera\electronvolt} scale. The three main production processes are pair production, s-channel resonances and vector boson fusion (VBF)~\cite{delahaye:SNOWMASS}. VBF is useful at high energies as the cross-section grows logarithmically as $\sigma \approx \ln{(\frac{s}{M_X^2})}$, once the $\sqrt{s} = m_X$ theshold has passed. An example Feynman diagram of VBF is shown in Figure \ref{fig:boson}. These large rates would allow for muon colliders to act as an electroweak boson collider at high energies.

A high energy and high resolution machine would be ideal for extending the Higgs sector, using Higgs self coupling and tri-Higgs production, as these are rare processes. This would allow measurements of the shape of the Higgs potential at higher orders. Studies suggest that a \SI{3}{\tera\electronvolt} muon collider has higher sensitivity and cross-section to Higgs self-coupling than CLIC, but lower statistics due to luminosity~\cite{Conway:Higgsself}. \\

Even though MAP demonstrates that a \SI{3}{\tera\electronvolt} muon collider provides strong physics potential, the lack of new particles discovered at LHC's energy frontier is a problem for motivating multi-TeV experiments. Should this change with new physics discoveries at the LHC upgrades, then a dedicated high-luminosity multi-TeV accelerator would be required to study such phenomena in more depth. It is theoretically expected for low-mass supersymmetrical particles to occur at around the TeV energy scale.

\subsection{Motivation summary}
To summarise, the muon's fundamentality, heavy mass and low energy spread allows for a muon collider machine to be used for both the precision frontier and the high energy frontier, therefore can both discover new signals and also make better precision measurements of known properties to test against the standard model predictions. This is a big advantage over current collider candidates and reduces the need for multiple next-generation colliders to provide for either precision or high energy, such as the FCC-ee upgrading to FCC-hh.\\

When compared to other next-generation accelerators, muon collisions can maximise energy production compared to proton collisions, have cleaner collisions than hadronic colliders, are not limited at high energies compared to circular electron colliders and have a better luminosity-power efficiency than electron colliders.\\

The muon collider provides an optimal environment for studying the Higgs resonance, due to its strong mass coupling, direct s-channel production and high energy resolution, however if other proposed colliders are able to study this resonance sooner, then there is less demand for a Higgs-factory muon collider to be built by 2050. \\

Whilst the muon collider shows promise and has a competitive physics reach, the feasibility of the muon collider decreases if no new justification for high energy physics is discovered at the LHC in the coming years. If the ESG judges that fields such as flavour physics and CP violation are the strongest methods of discovering new physics, then efforts will not be directed towards the muon colliders or other energy frontier machines.

\section{Technology feasibility}
\label{sec:Collider}
\begin{figure}[h]
    \centering
    \includegraphics[width=0.8\textwidth]{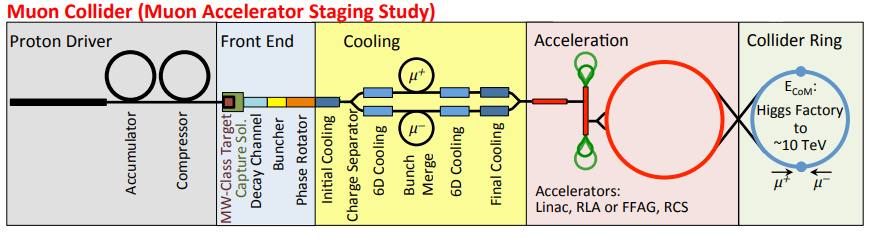}
    \caption{Block Schematic of Proton-Drive Muon Collider Concept, produced by MAP~\cite{Delahaye:Strategy}.}
    \label{fig:MAP}
\end{figure}

A muon collider is not a single asset, but instead requires multiple components which must function as a whole. To judge whether a muon collider will be feasible in 30 years the technological availability of each component must be considered. The conventional design for the muon collider uses a proton-driven source, a schematic of which is shown in Figure \ref{fig:MAP}. There are five crucial stages, all of which have to be experimentally tested and verified as individual components and then function coherently when assembled together. \\

The Muon Accelerator Program at Fermilab ran from 2010-2014 with the aim of producing feasibility studies for a muon collider by summarising the technologies to ensure that there is nothing that could inhibit production~\cite{Bross:MAP}. Whilst funding for this program was cut early, some of the key technological issues they have highlighted are discussed in this section to understand if they can be achieved in 30 years, focusing on the source, muon cooling and the acceleration techniques.

\subsection{Types of designs}
When discussing the building of a muon collider, it must be understood what design is being referred to. As part of the technology feasibility study, the MAP program has proposed multiple concepts with parameters displayed in Table \ref{tab:colliders}, which includes two factories at specific particle resonances, and four multi-TeV proposals for high energy discoveries. Section \ref{sec:MCType} will review these designs based on its physics benefits and technology requirements and conclude which one is most feasible to be operational in 30 years. \\

\begin{table}[!h]
\centering
\begin{tabular}{ccrrrrrr}  
\toprule
Parameter & & Higgs & Top & \SI{1.5}{\tera\electronvolt} & \SI{3}{\tera\electronvolt} & \SI{6}{\tera\electronvolt} & \SI{14}{\tera\electronvolt}\\
\midrule
CoM Energy & \si{\tera\electronvolt}&  0.126 & 0.35 & 1.5 & 3 & 6 & 14\\
Av. Luminosity & \SI{E34}{\per\square\centi\meter\per\second} & 0.008 & 0.6 & 1.25 & 4.4 & 12 & 40\\
Frequency & \si{\hertz} & 15 & 15 & 15 & 12 & 6 & 3.7 \\
Beam Power & \si{\mega\watt} & -& -& 6.75 & 10.8 & 10.8 & 15.4\\
Circumference & \si{\kilo\meter} & 0.3 & 0.7 & 2.5 & 4.5  & 6 & -\\
\bottomrule
\end{tabular}
\caption{Key collider parameters for 6 types of muon collider designs. \cite{Delahaye:Strategy}, \cite{Schulte:workshop}}
\label{tab:colliders}
\end{table}

\subsection{Key Technologies}
\subsubsection{Source}
\label{sec:source}
\subsubsection*{Proton source}
The proton driver is where protons are made, accelerated, and then collided into a target to produce pions which then decay to muons. This requires a standard \SI{3}{\giga\electronvolt} superconducting proton linac. Other technologies involved in the proton driver, such as the hydrogen ion stripping and final focusing magnet, will need to be demonstrated to prove they can perform as required, although this is not expected to be challenging. \\
After this, the front end deals with the target and organises the beam. The Higgs factory design requires a high powered \SI{4}{\mega\watt} target, which is technically difficult, as studied at MERIT, a mercury target tested at the proton synchrotron~\cite{McDonald:MERIT}. The remainder of the muon collider designs only require  1-2 \si{\mega\watt}, which can be achieved using a carbon target. There should be flexibility to allow for an upgrade to \SI{4}{\mega\watt} if required.

\subsubsection*{Positron source}
Instead of producing muons from protons, instead annihilating electron-positron pairs at the \Pmuon \APmuon threshold would create a high energy, low-emittance muon beam, which would have a longer lifetime and not require further cooling~\cite{Allport:positron}. This method is currently being investigated by the Low EMittance Muon Accelerator, (LEMMA) collaboration~\cite{Alesini:LEMMA}.

In order for the LEMMA beam to match \num{1E12} muons per bunch, a high intensity, high power positron beam is required. This power could be as great as \SI{1}{\tera\watt}, which is significant enough to need to be recycled, rather than a single use beam which is dumped. This would require an energy-recovering linac, which do not currently exist above the \si{\kilo\watt} scale. The intensity is estimated to require 1000 bunches of \SI{5E11}{\per\second} positrons, which is more ambitious than current state-of-the-art positron sources.

In addition, current simulations for the LEMMA muon beam have a large energy spread, which would reduce the resolution of resonance peaks, and decrease the physics motivation for being used for precision or for measuring the Higgs resonance. \\

If the positron source option is to be considered for a muon collider within 30 years, there should be a clear test facility which can prove that a LEMMA design can produce the same quality of beam as a proton-driven design.

\subsubsection{Cooling}
\label{sec:cooling}
Particle beam cooling is the process of reducing the emittance of the beam and is essential if it is composed of secondary particles. Emittance is a measure of area of the beam in position and momentum phase-space, and must be small at the collision point to optimise luminosity. As proton-source muon beams are produced from decaying protons then decaying pions, they have a very large initial emittance, and require a \num{E6} reduction in 6D emittance to reach the luminosities required for a Higgs factory (\SI{E32}{\per\centi\meter\squared\per\second}) or a TeV muon collider (\SI{E34}{\per\square\centi\meter\per\second}). If there is strong experimental evidence that the LEMMA positron source will be successful, then this negates the necessity for muon cooling, as mentioned in Section \ref{sec:source}.\\

Beam cooling is particularly challenging for muon beams as standard cooling techniques in a synchrotron, such as stochastic cooling, are not fast enough to occur on the required timescale. The proposed method for muon cooling is ionisation cooling, which involves the muon beam passing through a series of magnetic lattices and energy absorbers. This acts to minimize Coulomb-scattering within the material using strong focusing and a low-Z absorber, such as hydrogen. Cooling is optimised at the minimum ionisation energy, which in this case is \SI{200}{\mega\electronvolt}.\\

The Muon Ionisation Cooling Experiment (MICE) is an experiment which ran from 2001 - 2010 to demonstrate 4D ionisation cooling of a muon beam.~\cite{D.Kaplan:MCC_MICE}. It was successful in measuring reduced transverse amplitude of muons after liquid hydrogen and lithium hydride as absorbers~\cite{blackmore:MICE}. Despite this, there were limitations to this experiment: The muon beam was of low intensity with only \num{3.5E8} total events, MICE only tested one magnetic lattice cell, and less than 10\% emittance reduction was seen, which is not sigificant enough to base a collider on it. Furthermore, as this was only a 4D experiment, it did not attempt cooling in the longitudinal beam direction to make it 6D. \\

It is crucial for significant cooling to be demonstrated before further muon collider plans can go ahead. To prove that muon colliders are technologically viable, a full 6D cooling experiment at with an \num{E6} reduction in emittance is essential. There are currently functioning design simulations for 6D cooling, but there are no plans to develop a 6D cooling demonstration experiment~\cite{D.Kaplan:MCC_MICE}. Without muon cooling, then a significant factor of \num{E3} in luminosity will be lost.  If attempts at full muon cooling are not successful, then LEMMA may be the only option for a muon collider. If neither of these technologies cannot be proven over the next few years, there is no foundation to ensure that muon colliders can be operational by 2050. 


\subsubsection{Accelerator}
\label{sec:accelerator}
Significant feasibility studies of muon beam acceleration methods have not been fully performed. The difficulty of the accelerator method and the size of the ring depends on the energy selected, but it is likely that a combination of linacs and synchrotrons will be chosen. \\

In standard synchrotron technology, the superconducting magnets ramp up and increase in gradient as the particles increase in energy with each turn. For multi-TeV muon acceleration, this cycling must occur at a very rapid rate, due to the muon's short lifetime. This is not possible with current magnet technologies. Furthermore, for \si{\tera\electronvolt} acceleration, high-temperature superconducting (HTS) magnets will likely be required, which are still developing technology which will take many more years of production, and cannot be relied upon to be available within the next 30 years. \\

\begin{figure}[h!]
    \centering
    \includegraphics[width=0.7\textwidth]{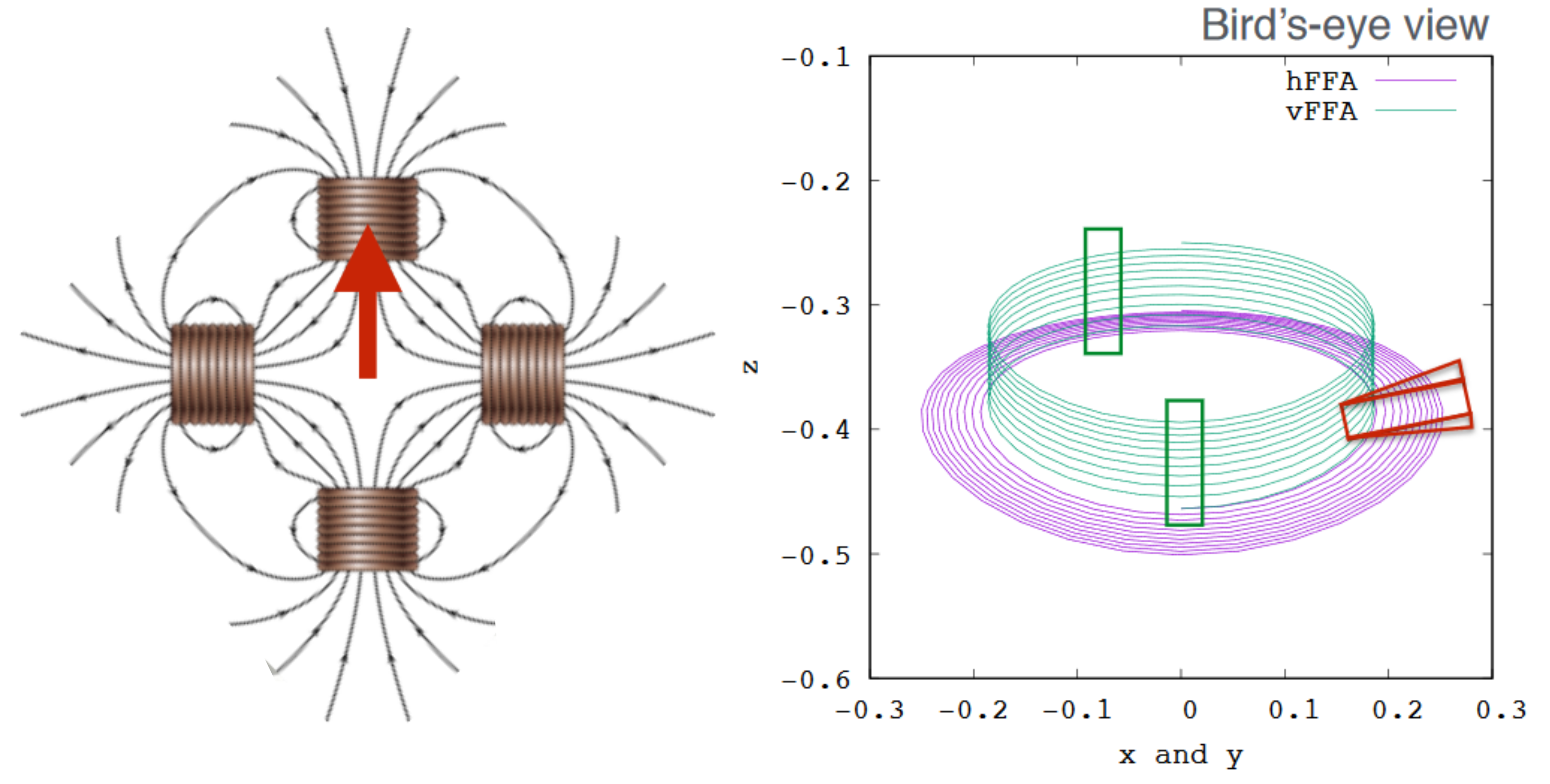}
    \caption{Field-lines of a vFFA magnet and orbit of an accelerating particle in a vFFA ring~\cite{Machida:FFA}.}
    \label{fig:FFA}
\end{figure}

An additional method for multi-TeV acceleration is to use vertical Fixed Field Alternating (vFFA) gradient magnets, which give a fixed field during acceleration, with a strong focusing effect. The magnet gradient would stay constant, but as the particles energy increased, the particle orbit would increase vertically, as shown in Figure \ref{fig:FFA}. vFFA magnets are being tested in the EMMA accelerator~\cite{BARLOW:EMMA} (Electron Model of Muon Acceleration, changed to Electron Model of Many Applications). It is possible that both rapid cycling and FFA magnets are used in conjunction in the accelerating ring. \\

To reduce costs and reuse current equipment, studies have suggested adapting the LHC tunnel (\SI{14}{\tera\electronvolt} ~\cite{Neuffer:LHCmumu}) or the FCC-ee ring (\SI{100}{\tera\electronvolt}~\cite{Zimmermann:FCCmumu}) for the purposes of muon colliders. Generally these studies have concluded that due to the space required for the cooling facility, re-using tunnels is only feasible for positron-sourced muons, which is dependant on the success of the LEMMA project.  \\

After the accelerating ring, the colliding ring has no significant technological difficulties, and therefore development of the optics lattice has already began as part of the MAP program, including for the \SI{3}{\tera\electronvolt} energy design~\cite{Alexahin:3tev}.

\subsubsection{Areas for further development}
\label{sec:detneut}
\subsubsection*{Detectors}
\label{sec:detectors}
Designing and producing a detector for colliders at this energy and intensity is well understood due to years of experience in the detector communities. In addition, the lower pile-up compared to hadronic collisions means it is less challenging to select and trigger on events. For this reason, ILC and CLIC detector simulations can be reused as placeholders for understanding the machine-detector interface. \\

Muons decaying to electrons will induce a strong background of electrons at the collision point, as mentioned in Section \ref{sec:challenges}. A proposed solution for this is to include a tungsten nozzle for shielding the detector from electrons to mitigate the background, with simulations suggesting that this does not significantly reduce the physics measurements~\cite{Richard:BriefBook}.\\

Before particles can collide within the detector, a series of focusing magnets is required to ensure good beam quality at the interaction point. When the circumference of the ring is small, such as \SI{0.3}{\kilo\meter} or \SI{0.7}{\kilo\meter} for the Higgs and Top Quark muon collider designs, the length of focusing magnets means that the collision ring can only support one interaction point. Having only one detector for an accelerator reduces the ability to cross-check any potential discoveries, and inhibits the reproducibility of the result.

\subsubsection*{Problems of neutrino radiation}
\label{sec:neutrino}
Effects of neutrino radiation on the TeV-scale can be mitigated with some techniques, such as undulating the beam, reducing the number of straight segments, halving emittance and having the collider deeper underground. Radiation significantly reduces for a LEMMA design, due to the smaller beam current. \\

FLUKA simulations of the neutrino radiation have been performed in preliminary studies which summarise that it becomes hazardous when a significant dose reaches the earth's surface, with a limit of \SI{0.1}{\milli\sievert} per year~\cite{bartosik:radiation}. These simulations have been performed up to a center-of-mass energy of \SI{3}{\tera\electronvolt} and suggest the distance required to be under \SI{0.1}{\milli\sievert} per year is approximately \SI{60}{\kilo\meter}, corresponding to a depth of \SI{250}{\meter}. The ESG highlights this problem and concludes that at energies above \SI{6}{\tera\electronvolt}, the luminosity of the muon beam may have to be reduced to reduce this radiation risk~\cite{Richard:BriefBook}. \\

A muon source could provide a high intensity neutrino beam for a neutrino detector. This could provide a mid-way stepping stone towards a muon collider, by making the source first. One potential neutrino factory being proposed using this method is called NuSTORM (Neutrinos from STORed Muons)~\cite{Long_2018}. This uses a decay ring with long lengths pointing in the direction of a detector, and can be used to study leptonic CP invariance violation and sterile neutrinos. The potential for this depends on the need for a new neutrino experiment, but with current work developing DUNE and Hyper-Kamiokande (\cite{Richard:BriefBook} Section 6: Neutrino Physics), this will not be a priority until current experiments have been finalised.

\subsubsection{Technology Summary}
This section briefly summarises the conclusions from technological feasibility to highlight any necessary significant developments.\\

Assuming a proton-driven source, the high energy target should be feasible up to 1-2 \si{\mega\watt}, any higher requires a dedicated test facility.\\Cooling has not been achieved sufficiently to give confidence to its availability and should not be assumed to function until a dedicated 6D cooling facility has been completed.\\More feasibility studies for high energy acceleration is required. It if is determined that HTS are necessary for multi-TeV acceleration, then this will delay the timescale of a muon collider beyond 30 years.

\subsection{Optimal design}
\label{sec:MCType}
This section concludes the optimal muon collider design for the short-term needs of the particle physics community by considering their respective physics reaches and technological limits.\\

For the top quark design, although the design is smaller and does not require the difficult methods of multi-TeV acceleration, the requirements for a top physics muon collider have decreased due to the success of the LHC as a ``top factory"~\cite{Husemann:Top}, and competitive future accelerators can continue this work. For this reason, it will not be considered as a priority design.\\

The Higgs factory is potentially the most technologically challenging muon collider, due to the high power \SI{4}{\mega\watt} mercury target that is required. It would require a dedicated facility to ensure feasibility, or for the upgrade of an existing 1-2 \si{\mega\watt} target. Both situations would require more time, and restricts the ability for the Higgs factory to act as a stepping-stone towards higher energy upgrades. In addition, it could only provide a single detector site, which limits the repeatability of the results and may cause need for a second Higgs factory to verify data. Currently there are multiple future lepton colliders being conceptualised which would act as a Higgs factory design to measure its resonance, and once one of them is built, the motivation for this design drops. For these reason, it will not be considered as a design to be completed by 2050.\\

The \SI{14}{\tera\electronvolt} design is less developed, as it has only recently been suggested as a rival for the physics reach of FCC-hh~\cite{Schulte:workshop}. Although \SI{6}{\tera\electronvolt} and \SI{14}{\tera\electronvolt} muon colliders have the best energy frontier reach, the high energy would put strain on demand for the acceleration requirements, which should be reserved for when the technology is well understood. It would be difficult to jump straight from R\&D to a 6 or 14 \si{\tera\electronvolt} collider within the 30 year limit, particularly with no specific high energy physics goal in mind.
Most importantly, to reduce the strong neutrino radiation, the luminosity of these beams would have to be reduced, and extra caution would have to be taken in the design and engineering of the collider, to protect people at the surface. These factors discount them from being viable for 2050.\\

Both the \SI{1.5}{\tera\electronvolt} and the \SI{3}{\tera\electronvolt} show potential in being able to provide new physics, but are small enough to be feasible to construct after the R\&D stage. The \SI{3}{\tera\electronvolt} will be taken as the most viable muon collider design, aiming to be operational within the next 30 years.

\section{Discussion}
\begin{figure}[h!]
    \centering
    \includegraphics[width=1\textwidth]{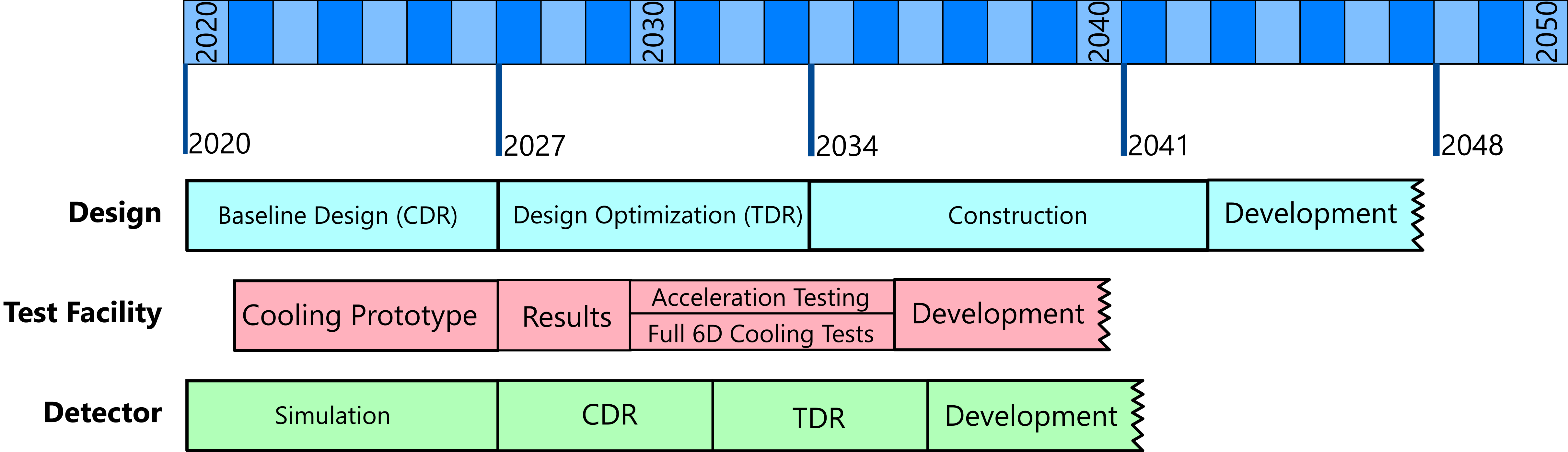}
    \caption{Proposed roadmap for a \SI{3}{\tera\electronvolt} muon collider, (inspired by \cite{Schulte:lumi}, slide 23)}
    \label{fig:roadmap}
\end{figure}

This report has summarised the benefits of a muon collider, the physics reach that it would provide, and the technological difficulties that need to be overcome in order for a muon collider to be operational within the next 30 years.\\
The evidence and conclusions from each section can be brought together to suggest a potential roadmap that would allow a muon collider to be operational by 2050, an outline of which is shown in Figure \ref{fig:roadmap}. It is important to align the project milestones with 7-yearly European Particle Physics Strategy Updates, to provide frequent deadlines.  \\

As per Section \ref{sec:cooling}, the significant limiting technological factor is to prove that a cooling system can reduce the 6D beam emittance by \num{E6}. The existence of a small-scale experimental test facility to demonstrate 6D cooling would benefit the muon collider community, acting as the first step from simulations to a \SI{3}{\tera\electronvolt} collider and the facility could host a working group for simultaneous development of hardware and simulations for the CDR.
The CDR should include a include a complete beginning-to-end simulation of the whole facility, to ensure that each individual module of Figure \ref{fig:MAP} can work coherently and provide a final beam that meets the requirements for the physics outlined in Section \ref{sec:physics}. Ideally, a green light from this year's ESG should allow for work on the cooling facility and the CDR to start immediately, so that prototype cooling results and the completed CDR can be finalised for the next particle physics strategy update in 2027. \\
From there, approval from the ESG would allow for development of the technical design report (TDR) to begin, and promote the further facilities to test multi-TeV muon acceleration methods mentioned in Section \ref{sec:accelerator}. If the TDR and the test facilities are developed by the 2034 update, then this would give 16 years devoted to developing, building and operating a full-scale \SI{3}{\tera\electronvolt} muon collider. This includes the time to construct the \SI{4.5}{\kilo\meter} circumference tunnel.

\section{Conclusion}
To answer the question of ``Can a muon collider be operational within the next 30 years?", this critique has presented the motivations and physics benefits of muon colliders, summarised the core technologies required, selected a design option from those produced, and then proposed a roadmap to begin developing the chosen muon collider design. \\

This outline plan is optimistic, and assumes that funding will be available immediately to be used on small-scale R\&D projects within the next few years. In addition, it assumes that no other competitive high-energy collider design will be chosen in place of the muon collider, and that the European Particle Physics Strategy Group will continue to support and encourage its construction. In addition, the overall plan requires many physicists and engineers working on this full-time, and requires the simulation and hardware testing to be occurring simultaneously.\\

Should all of these requirements be fulfilled, then a muon collider can be made operational by 2050.\\

In reality, the muon collider collaboration will be competing with more developed accelerator concepts for funding, which already have a full-time work force and functioning test facilities. To be considered a priority, the muon collider must demonstrate greater high energy potential in its physics discoveries to gain the full support of the ESG in 2020. As this is unlikely to occur, muon colliders may be considered advanced accelerator technology, alongside similar ambitious projects such as plasma wakefield accelerators and high temperature superconductors~\cite{ECFA:2019}. These future technologies are expected to function in the era of 2060-2080, therefore not considered to operate within the next 30 years. \\

All concepts and conclusions are due to the current high energy physics epoch that has followed since the discovery of the Higgs Boson. Should the LHC discover new and unexpected physics within the remainder of its lifetime, then current accelerator designs will certainly adapt and optimise to suit the discovery. In such an event, the focus will shift from a Higgs-factory style design into a discovery machine, to precisely find new particles at high energies, and then the muon collider would become the ideal candidate.

\printbibliography

@article{Richard:BriefBook,
      author        = "Ellis, R and Heinemann, B",
      title         = {Physics Briefing Book: Input for the European Strategy for Particle Physics Update 2020},
      note        = "CERN-ESU-004",
      year          = "2019"
}

@article{Delahaye:Strategy,
    title={Muon Colliders: Input to the European Particle Physics Strategy Update},
    author={Delahaye, J and Diemoz, M},
    year={2019},
    eprint={1901.06150},
    archivePrefix={arXiv}
}

@article{Delahaye:staged,
    title={A Staged Muon Accelerator Facility For Neutrino and Collider Physics},
    author={Delahaye, J and Ankenbrandt, C},
    year={2015},
    eprint={1502.01647},
    archivePrefix={arXiv}
}

@article{Tanabashi:PDG,
  title = {Review of Particle Physics},
  author = {Tanabashi, M and others},
  collaboration = {Particle Data Group},
  journal = {Phys. Rev. D},
  volume = {98},
  issue = {3},
  pages = {030001},
  numpages = {1898},
  year = {2018},
  publisher = {American Physical Society},
  doi = {10.1103/PhysRevD.98.030001}
}

@article{Boscolo:FutureMuon,
      author         = "Boscolo, M and Delahaye, J and Palmer, M",
      title          = "{The future prospects of muon colliders and neutrino factories}",
      journal        = "Rev. Accel. Sci. Tech.",
      volume         = "10",
      year           = "2019",
      number         = "01",
      pages          = "189-214",
      doi            = "10.1142/9789811209604_0010, 10.1142/S179362681930010X",
      eprint         = "1808.01858",
      archivePrefix  = "arXiv"
}

@article{D.Kaplan:MCC_MICE,
author = {D. Kaplan},
title = {Muon colliders, neutrino factories, and results from the MICE experiment},
journal = {AIP Conference Proceedings},
volume = {2160},
number = {1},
pages = {040011},
year = {2019},
doi = {10.1063/1.5127691},
}

@article{BARLOW:EMMA,
title = "EMMA—The world’s first non-scaling FFAG",
journal = "Nuclear Instruments and Methods in Physics Research Section A: Accelerators, Spectrometers, Detectors and Associated Equipment",
volume = "624",
number = "1",
pages = "1 - 19",
year = "2010",
doi = "https://doi.org/10.1016/j.nima.2010.08.109",
author = "R. Barlow and others",
}

@article{rubbia:Higgs,
    title={A complete demonstrator of a muon cooled Higgs factory},
    author={Rubbia, C},
    year={2013},
    eprint={1308.6612},
    archivePrefix={arXiv}
}

@article{Palmer:Higgs,
    title={Muon Colliders: Physics and Accelerator Technology},
    author={Palmer, M.},
    journal={Higgs-Maxwell Particle Physics Workshop Royal Society of Edinburgh},
    year={2016},
    pages={9}
}

@article{Schulte:lumi,
    title={Muon Collider: A path to the future?},
    author={Schulte, D},
    journal={European Physical Society Conference on High Energy Physics},
    year={2019},
    pages={32-36},
    url={https://indico.cern.ch/event/577856/contributions/3420383/}
}

@article{Schulte:workshop,
    title={Muon Collider Strategy},
    author={Schulte, D},
    journal={Muon Collider Workshop, CERN},
    year={2019},
    pages={19},
    url={https://indico.cern.ch/event/845054/contributions/3573351/}
}

@article{delahaye:SNOWMASS,
    title={Enabling Intensity and Energy Frontier Science with a Muon Accelerator Facility in the U.S.},
    author={Delahaye, J and Ankenbrandt, C},
    year={2013},
    journal={U.S. Community Summer Study of the Division of Particles and Fields of the American Physical Society},
    eprint={1308.0494},
    archivePrefix={arXiv}
}

@article{Husemann:Top,
   title={Top-quark physics: Status and prospects},
   volume={95},
   DOI={10.1016/j.ppnp.2017.03.002},
   journal={Progress in Particle and Nuclear Physics},
   publisher={Elsevier BV},
   author={Husemann, Ulrich},
   year={2017},
   pages={48–97}
}

@article{Bross:MAP,
author = {Bross,A},
title = {The US Muon Accelerator Program (MAP)},
journal = {AIP Conference Proceedings},
volume = {1382},
number = {1},
pages = {59-63},
year = {2011},
doi = {10.1063/1.3644270}
}

@article{Alesini:LEMMA,
    title={Positron driven muon source for a muon collider},
    author={Alesini, D and Antonelli, M and others},
    year={2019},
    eprint={1905.05747},
    archivePrefix={arXiv}
}

@article{Barger:physics1998,
   title={Physics at muon colliders},
   DOI={10.1063/1.55093},
   journal={American Institute of Physics},
   author={Barger, V.},
   year={1998}
}

@article{bartosik:radiation,
    title={Preliminary Report on the Study of Beam-Induced Background Effects at a Muon Collider},
    author={Bartosik, N and Bertolin, A},
    year= {2019} ,
    eprint={1905.03725},
    archivePrefix={arXiv}
}

@conference{ECFA:2019,
    author    = "D'Hondt, J",
    title     = "Open Session on Advanced Accelerator Technologies",
    booktitle = "105th Plenary ECFA meeting",
    year      = "2019"
}

@article{McDonald:MERIT,
      author         = "McDonald, K and Kirk, H and others",
      title          = "{The MERIT High-Power Target Experiment at the CERN PS}",
      journal      = "{Proceedings of the 23rd Particle Accelerator Conference, PAC'09}",
      year           = "2010"
}

@article{Allport:positron,
    title={Producing an Intense, Cool Muon Beam via e+e- Annihilation},
    author={Kaplan, D and Allport, P},
    year={2007},
    eprint={0707.1546},
    archivePrefix={arXiv},
    note={NuFact06}
}

@article{blackmore:MICE,
    title={Recent results from the study of emittance evolution in MICE},
    author={Blackmore, V. and MICE Collaboration},
    year={2018},
    eprint={1806.04409},
    archivePrefix={arXiv}
}

@article{Alexahin:3tev,
author = {Alexahin, Y. and Gianfelice-Wendt, E.},
year = {2012},
journal= {FERMILAB-CONF-12-184-APC},
title = {A 3 TeV Muon Collider Lattice Design}
}

@article{Long_2018,
	doi = {10.1088/1742-6596/1056/1/012033},
	year = 2018,
	publisher = {{IOP} Publishing},
	volume = {1056},
	pages = {012033},
	author = {Long, K},
	title = {The {nuSTORM} experiment},
	journal = {Journal of Physics: Conference Series}
}

@article{Neuffer:LHCmumu,
	doi = {10.1088/1748-0221/13/10/t10003},
	year = 2018,
	publisher = {{IOP} Publishing},
	volume = {13},
	number = {10},
	pages = {T10003--T10003},
	author = {Neuffer, D and Shiltsev, V},
	title = {On the feasibility of a pulsed 14 {TeV} c.m.e. muon collider in the {LHC} tunnel},
	journal = {Journal of Instrumentation}
}

@article{Zimmermann:FCCmumu,
	doi = {10.1088/1742-6596/1067/2/022017},
	year = 2018,
	publisher = {{IOP} Publishing},
	volume = {1067},
	pages = {022017},
	author = {F. Zimmermann},
	title = {{LHC}/{FCC}-based muon colliders},
	journal = {Journal of Physics: Conference Series}
}

@article{Conway:Higgsself,
      author         = "Conway, Alexander and Wenzel, Hans and others",
      title          = "{Measuring the Higgs Self-Coupling Constant at a
                        Multi-TeV Muon Collider}",
      year           = "2014",
      eprint         = "1405.5910",
      archivePrefix  = "arXiv"
}

@conference{Machida:FFA,
    author    = "Machida, S",
    title     = "ISIS upgrade and the feasibility study",
    booktitle = "International Workshop on Fixed Field alternating gradient Accelerators",
    slide    = "10",
    year      = "2019"
}

\end{document}